\documentclass[a4paper,12pt,english]{article}
\usepackage{amssymb,bbm,babel,graphicx,multicol}
\usepackage{psfrag}
\usepackage{color}
\usepackage{hyperref}

\makeatletter
\@addtoreset{equation}{section}
\makeatother

\def\dalemb#1#2{{\vbox{\hrule height .#2pt
        \hbox{\vrule width.#2pt height#1pt \kern#1pt
                \vrule width.#2pt}
        \hrule height.#2pt}}}

\def\cA{{\cal A}}

\def\cD{{\cal D}}

\def\hp{ \frac{1}{2}}
\def\qu{\fr{1}{4}}

\let\a=\alpha \let\b=\beta \let\g=\gamma \let\d=\delta \let\e=\epsilon
  \let\q=\theta  
\let\l=\lambda \let\m=\mu \let\n=\nu  \let\r=\rho
 \let\t=\tau

 \let\W=\Omega

\def\nn{\nonumber} \def\bd{\begin{document}} \def\ed{\end{document}}
\def\ds{\documentstyle} \let\fr=\frac \let\bl=\bigl \let\br=\bigr
\let\Br=\Bigr \let\Bl=\Bigl
\let\bm=\bibitem
\let\na=\nabla
\let\pa=\partial \let\ov=\overline
\let\ul=\underline
\newcommand{\be}{\begin{equation}}
\newcommand{\ee}{\end{equation}}
\def\ba{\begin{array}}
\def\ea{\end{array}}
\def\ft#1#2{{\textstyle{{\scriptstyle #1}\over {\scriptstyle #2}}}}
\def\fft#1#2{{#1 \over #2}}
\def\del{\partial}
\def\sst#1{{\scriptscriptstyle #1}}
\def\oneone{\rlap 1\mkern4mu{\rm l}}
\def\ie{{\it i.e.\ }}
\def\via{{\it via}}
\def\semi{{\ltimes}}
\def\str{{\rm str}}
\def\Dm{{{D_{\sst{max}}}}}
\def\vac{ \left | 0 \right \rangle }
\def\kvac{ \left | k \right \rangle }

\def\sp{\; \; \;}

\def\bol{ \left | B (p^+) \right \rangle}
\def\bo1{ \left | B^0 (p^+) \right \rangle}

\def\bolt{ \left | B (p^+) \right \rangle_{\t}}
\def\boxl{ \left | B (x^-) \right \rangle}

\def\<{ \langle }
\def\>{ \rangle }

\def\vf{\varphi}

\def\ls{{(l,0)}}
\def\lv{{(l,\pm1)}}
\def\lt{{(l,\pm2)}}

\def\lse#1{{(l_{#1},0)}}
\def\lve#1{{(l_{#1},\pm1)}}
\def\lte#1{{(l_{#1},\pm2)}}

\def\lsg#1{{5(l_{#1},0)}}
\def\lvg#1{{5(l_{#1},\pm1)}}
\def\ltg#1{{5(l_{#1},\pm2)}}

\def\lsi#1{{5{(#1,0)}}}
\def\lvi#1{{5{(#1,\pm1)}}}
\def\lti#1{{5{(#1,\pm2)}}}

\def\lsr#1{{1{(#1,0)}}}
\def\lvr#1{{1{(#1,\pm1)}}}
\def\ltr#1{{1{(#1,\pm2)}}}

\def\cn{{\cal N}}
\def\cao{{\cal O}}
\def\cD{{\cal D}}
\def\cE{{\cal E}}
\def\cF{{\cal F}}
\def\cG{{\cal G}}
\def\cH{{\cal H}}
\def\cK{{\cal K}}
\def\cO{{\cal O}}
\def\cP{{\cal P}}
\def\cQ{{\cal Q}}
\def\cR{{\cal R}}
\def\cS{{\cal S}}
\def\cT{{\cal T}}
\def\cU{{\cal U}}
\def\cV{{\cal V}}
\def\cW{{\cal W}}

\newcommand{\nono}{\nonumber}
\newcommand{\eqref}[1]{(\ref{#1})}
\newcommand{\dtilde}[1]{\tilde{\tilde{#1}}}
\newcommand{\hatb}[1]{\hat{\ov{#1}}}
\newcommand{\hatt}[1]{\hat{\tilde{#1}}}
\newcommand{\emnr}{{e_\m}^{\n\r}}
\newcommand{\psm}{\int \frac{d^{11}\l}{{\l^+}^3}}

\newcommand{\hsp}{\hspace{0.5cm}}

\newcommand{\ho}[1]{$\, ^{#1}$}
\newcommand{\hoch}[1]{$\, ^{#1}$}
\newcommand{\bea}{\begin{eqnarray}}
\newcommand{\eea}{\end{eqnarray}}
\newcommand{\ra}{\rightarrow}
\newcommand{\lra}{\longrightarrow}
\newcommand{\Lra}{\Leftrightarrow}
\newcommand{\lera}{\leftrightarrow}
\newcommand{\ap}{\alpha^\prime}
\newcommand{\bp}{\tilde \beta^\prime}
\newcommand{\tr}{{\rm tr} }
\newcommand{\Tr}{{\rm Tr} }
\newcommand{\NP}{Nucl. Phys. }

\newcommand{\ams}
{${}^1$ {\it Instituto de Fisica Teorica, State University of Sao Paulo, \\
Rua Dr. Bento T. Ferraz 271, Sao Paulo, SP 01140-070, Brasil\\
{\tt nberkovi@ift.unesp.br}} \vspace{0.5 cm} \\
{\it ${}^2$Institute for Theoretical Physics,
University of Amsterdam, \\
Valckenierstraat 65, 1018XE Amsterdam, The Netherlands} \\
{\tt J.Hoogeveen, K.Skenderis@uva.nl}  \vspace{0.5 cm} \\
{${}^3$ {\it Kavli Institute for Theoretical Physics\\
University of California at Santa Barbara,\\
Santa Barbara, CA 93106-4030, USA}}}

\newcommand{\auth}{Nathan Berkovits${}^{1,3}$, Joost Hoogeveen${}^2$ and Kostas Skenderis${}^{2,3}$}

\def\red{\color{red}}

\vspace{10pt}

\begin{document}

\begin{flushright}
\hfill{ITF-2009-16}
\\
\hfill{NSF-KITP-09-103}
\\
\hfill{IFT-P.005/2009}

\end{flushright}

\vspace{25pt}

\begin{center}

  {\Large \bf Decoupling of unphysical states in the minimal pure spinor formalism II}

\vspace{20pt}

\auth

\vspace{15pt}

\vspace{8pt}

{\ams}

\vspace{20pt}

\underline{ABSTRACT}

\end{center}

\noindent This is the second of a series of two papers where decoupling of unphysical states in the minimal pure spinor formalism is investigated. The multi-loop amplitude prescription for the minimal pure spinor superstring formulated in hep-th/0406055 involves the insertion of picture changing operators in the path integral. In the first paper it was shown that these operators are not BRST closed inside correlators. Therefore a new proof of decoupling of BRST exact states is needed. In this paper we present such a proof, which applies to arbitrary genus. It relies in part on a (previously unnoticed) invariance of the path integral measure.

\pagebreak



\section{Introduction}
A new superstring formalism, the pure spinor formalism, has been developed over the past ten years \cite{Berkovits:2000fe,Berkovits:2004px,Berkovits:2005bt}, see \cite{Berkovits:2002zk, Mafra:2009wq} for reviews. In this new formalism, the theory exhibits manifest super Poincar\'{e} invariance,
as in the Green-Schwarz (GS) formalism, but in contrast
with the GS string the worldsheet theory in flat target space is
free, as in the Ramond-Neveu-Schwarz (RNS) formalism,
so the theory can be quantized straightforwardly.

There are two versions of this formalism, the minimal \cite{Berkovits:2004px}
and the non-minimal formalism \cite{Berkovits:2005bt}.
In this paper we discuss exclusively the minimal case.
A multi-loop scattering amplitude prescription was developed in
\cite{Berkovits:2004px} and it involves introducing a number of
picture changing operators (PCO's) that are inserted in the path integral.
These operators are BRST closed in a distributional sense
and depend on constant spinors
($C_{\a}$) and constant tensors ($B_{mn}$).
It was argued in \cite{Berkovits:2004px}
that amplitudes are independent of $C$ and $B$, because the Lorentz
variation of PCO's is BRST exact. The same conclusion can be reached
from the results in \cite{Hoogeveen:2007tu}, where the presence of
the PCO's in the path integral was derived via BRST-BV quantization that
takes into account the gauge invariances due to zero modes
\cite{Craps:2005wk}. A manifestly Lorentz invariant prescription
can be obtained by integrating over $C$ and $B$ \cite{Berkovits:2004px}
and this is the form we will use here.

One may question whether the PCO's $Y$ are closed because their BRST variation only vanishes in a distributional sense,
$Q Y \sim x \delta(x)$, where $x$ is an appropriate field variable. That would give zero only if integrated against a smooth function of $x$. It is shown in \cite{new} however that the amplitudes in the current minimal amplitude prescription do contain singular terms that imply that $Q Y$ is not zero inside correlators.
This leads to explicit dependence of the amplitudes on $B$ and $C$ and
problems with decoupling of BRST exact states and with Lorentz
invariance, when one does not integrate over $B$ and $C$.
The singular behavior of the amplitudes is linked to the fact that the gauge fixing condition of the zero mode gauge invariances implicit in the current formulation is singular and we anticipate that a proper treatment of global
issues will allow for a non-singular gauge fixing condition and will lead to a prescription free of these
problems. Work in this direction is in progress \cite{progress}.
Here we will show that these problems
are also absent when one integrates over $B$ and $C$.

Note that $Q Y \neq 0$ by itself does not imply that $Q$ exact states do not decouple. It only
implies that the standard argument for decoupling of unphysical states that involves integrating
$Q$ by parts does not automatically lead to decoupling. In \cite{new} it was shown that at tree level
one can nevertheless establish decoupling of BRST exact states using integration of $Q$ by parts. The core of that argument is the vanishing of the trace of a certain Lorentz invariant tensor, $(\e T)$, defined in \eqref{eq:defet}. At one loop, we can identify a Lorentz invariant tensor $(\e T R)$, defined in \eqref{eq:defetR}, that is the one-loop analog of $(\e T)$. That is to say, all one-loop amplitudes with a $Q$ exact state are proportional to the trace of this tensor. Had this trace vanished, this would imply that $Q$ exact states decoupled at one loop. It turns out however, as proven in \cite{new}, that the trace of this one-loop invariant tensor does not vanish, so one needs a different argument to prove decoupling of unphysical states.

The main result of this paper is to provide such an argument. The new argument does not use
integration of $Q$ by parts. Rather it
makes use of a (so far unnoticed) invariance of the path integral measure and the fact the zero mode integrals act as projectors on a certain Lorentz scalar. Then one can show that the integrand that results
from BRST exact insertions
does not contain this scalar, hence amplitudes that contain unphysical states vanish after integration.

This paper is organized as follows. In the next section we present a short review of the amplitude prescription in the minimal pure spinor formalism, mostly to set our notation. In section \ref{sec:proof} the main result is discussed, namely the proof of decoupling of unphysical states in the minimal pure spinor formalism.  Finally there is a short concluding section.

\section{Minimal pure spinor amplitude prescription}\label{sec:rev}
In this section we review the amplitude prescription for the
minimal pure spinor superstring \cite{Berkovits:2004px}. The $N$ point tree-level amplitude is given by
\be
\cA=\< V_1(z_1)V_2(z_2)V_3(z_3)\int dz_4 U_4(z_4) \cdots \int dz_N U_N(z_N)Y_{C_1}(y_1)\cdots Y_{C_{11}}(y_{11})\>
\ee
The vertex operators are given by
\bea
V&=&\lambda^{\a}A_{\a}(x,\q),
\\
U&=&\del \q^{\a}A_{\a}(x,\q)+\Pi^mA_m(x,\q)+d_{\a}W^{\a}(x,\q)+\hp N^{mn}\mathcal{F}_{mn}(x,\q).
\eea
where $A_{\a}(x,\q), A_m(x,\q)$ is the superfield connection for
$N{=}1, d{=}10$\ SYM
and $W^{\b}, \mathcal{F}_{mn}$ are the corresponding field strengths. $Y_C$ are the picture changing operators (PCO) needed to absorb the zero modes of the weight zero fields:
\be
Y_C(y)=C_{\a}\q^{\a}(y)\d(C_{\b}\lambda^{\b}(y)),
\ee
where $C_{\a}$ is a constant spinor.

The $N$ point one-loop amplitude involves one unintegrated vertex operator,
$V$, $N{-}1$ integrated vertex operators, $U$, a composite $\tilde{b}$ field
and picture changing operators $Y_C, Z_B$ and  $Z_J$,
\be \label{eq:1looppre}
\cA^{(N)}=\int d^2 \t \< | \int d^2 u \m(u) \tilde{b}_{B^1}(u,z_1) \prod_{P=2}^{10}Z_{B^P}(z_P)Z_J(z_{11})\prod_{I=1}^{11}Y_{C_I}(y)|^2
\ee
\[
V_1(t_1)\prod_{T=2}^{N}\int d^2t_TU_T(t_T) \>.
\]


The $Z_B$'s and $Z_J$ are PCO's needed to absorb the zero modes of the weight one fields:
\be
Z_B(z)=\hp B_{mn}\lambda(z)\g^{mn}d(z)\d(B_{mn}N^{mn}(z)),\quad
Z_J(z)=\lambda^{\a}(z)d_{\a}(z) \d(J(z)),
\ee
where $B_{mn}$ is a constant antisymmetric tensor. The composite $\tilde{b}$ field satisfies
\be \label{eq:qbbuz}
\{Q,\tilde{b}_B(u,z)\}=T(u)Z_B(z).
\ee
This equation ensures the $Q$ variation of the $b$ ghost vanishes after integrating over moduli space. The solution is given by  \cite{Berkovits:2004px}
\be
\tilde{b}_{B}(u,z)=b_{B}(u)+T(u)\int_u^z dv B_{pq}\del N^{pq}(v)\d(BN(v)).
\ee
The local $b$ ghost, $b_B(u)$, is a composite operator, constructed out of the worldsheet fields:
\be \label{eq:bghost}
b_B(z)={b_B}_0(z)\d(BN(z))+{b_B}_1(z)\d'(BN(z))+{b_B}_2(z)\d''(BN(z))+{b_B}_3(z)\d'''(BN(z)),
\ee
where the primes denote derivatives, $BN\equiv B_{mn}N^{mn}$ and the explicit expressions of ${b_B}_i$ can be found in appendix \ref{sec:appco}.

The amplitude prescription for $g>1$ is given by
\be \label{eq:hlooppre}
\cA^{(N)}=\int d^2 \t_1\cdots d^2\t_{3g-3} \< | \prod_{P=1}^{3g-3}\int d^2 u_P \m_P(u_P) \tilde{b}_{B^P}(u_P,z_P)
\ee
\[
\prod_{P=3g-2}^{10g}Z_{B^P}(z_P) \prod_{R=1}^g
Z_{J^R}(v_R)\prod_{I=1}^{11}Y_{C_I}(y)|^2\prod_{T=1}^{N}\int d^2t_TU_T(t_T) \>.
\]
where we have grouped the insertions in a way that will be useful later.

The amplitude prescription just described was also obtained from first principles in \cite{Hoogeveen:2007tu} by coupling the pure spinor sigma model to
topological gravity. In particular, the PCO's arise in this
context by gauge fixing gauge invariances due to zero modes.
The constant tensors $C_\a$ and $B_{mn}$ enter the theory through
gauge fixing conditions. Thus, provided there are no BRST anomalies
and the gauge fixing is non-singular, the amplitudes should be
independent of $C$ and $B$. However, as shown in \cite{new}, the gauge fixing is singular and amplitudes do depend on the choice of $B$ and $C$. One can restore manisfest Lorentz invariance by integrating over all possible choices for $B$ and $C$. This integral is incorporated in the computations below.

As described in \cite{Berkovits:2004px},
the amplitude \eqref{eq:1looppre} is evaluated by first using the OPE's
to remove all fields of non-zero weight. After this step all fields
have weight zero. This can be evaluated by replacing the fields by
their zero modes and performing the zero mode integrations. So
we need to know how to integrate over the zero modes.
For the $d,\q,x$ variables this is standard, so we only discuss
the integration over $\lambda, N, B, C$.

A typical integral one encounters is given by \cite{Berkovits:2004px}:
\be \label{eq:cA}
\cA=\int [d\lambda][dB][dC]\prod_{R=1}^g[dN_R] f(\lambda,N_R,J_R,C,B)
\ee
where the zero mode measure for $[d\lambda]$ is given by
\be \label{eq:measlam}
[d\lambda]\lambda^{\a}\lambda^{\b}\lambda^{\g}=d\lambda^{\a_1}\wedge \cdots \wedge d\lambda^{\a_{11}} ( \e T)^{\a\b\g}_{\a_1\cdots \a_{11}},
\ee
with
\be \label{eq:defet}
(\e T)^{\a\b\g}_{\a_{1}\cdots \a_{11}}= \e_{\a_1\cdots \a_{16}}T^{\a\b\g\a_{12}\cdots\a_{16}},
\ee
\be \label{eq:deft}
T^{\a\b\g\a_{12}\cdots \a_{16}}=\g_m^{\a[\a_{12}}\g_n^{|\b|\a_{13}}\g_p^{|\g|\a_{14}}(\g^{mnp})^{\a_{15}\a_{16}]}.
\ee
The zero mode measure for (each zero mode of) $[dN]$ is given by
\be
[dN]\lambda^{\a_1}\cdots \lambda^{\a_8}=dN^{m_1n_1}\wedge \cdots \wedge N^{m_{10}n_{10}}\wedge dJ R^{\a_1\cdots \a_{8}}_{m_1n_1\cdots m_{10}n_{10}},
\ee
with
\be \label{eq:defR}
R^{\a_1\cdots \a_8}_{m_1n_1\cdots m_{10}n_{10}}\equiv \g^{((\a_1\a_2}_{m_1n_1m_2m_3m_4} \g^{\a_3\a_4}_{m_5n_5n_2m_6m_7} \g^{\a_5\a_6}_{m_8n_8n_3n_6m_9} \g^{\a_7\a_8))}_{m_{10}n_{10}n_4n_7n_9}+ {\rm permutations}.
\ee
The permutations make $R$ antisymmetric under exchange in both $m_i \lera n_i$ and $m_in_i \lera m_jn_j$ and the double brackets denote subtraction of the gamma trace. The zero mode integral \eqref{eq:cA} is only non-zero if the function $f$ depends on $(\lambda,N,J,C,B)$ as
\be \label{eq:fb}
f(\lambda,N,J,C,B)=
\ee
\[
h(\lambda,N,J,C,B)\prod_{R=1}^g\del^{M_R} \d(J)\prod_{P=1}^{10}\prod_{R=1}^g \del^{L_{P,R}}\d(B^P N_R) \prod_{I=1}^{11} \del^{K_I}\d(C^I \lambda),
\]
where the polynomial $h$ assumes the form
\be \label{eq:hb}
(\lambda)^{8g-8+\sum_{I=1}^{11}(K_I+1)}\prod_{R=1}^g(J_R)^{M_R} (N_R)^{\sum_{P=1}^{10} L_{P,R}} \prod_{P=1}^{10} (B^P)^{L_{P,R+1}}\prod_{I=1}^{11} (C^I)^{K_I+1}.
\ee
The integration over the zero modes of the pure spinor variables and the constant tensors is defined in \cite{Berkovits:2004px} as
\be\label{eq:bint}
\cA=c \fr{\del}{\del \lambda^{\a_1}} \cdots \fr{\del}{\del \lambda^{\a_{3}}} (\e T)^{\a_1\cdots \a_{3}}_{\b_1\cdots \b_{11}} \left[R^{\a_4\cdots \a_{11}}_{m_1n_1\cdots m_{10}n_{10}} \fr{\del}{\del \lambda^{\a_4}} \cdots \fr{\del}{\del \lambda^{\a_{11}}}  \fr{\del}{\del B^1_{m_1n_1}}\cdots \fr{\del}{\del B^{10}_{m_{10}n_{10}}}\right]^g
\ee
\[
\fr{\del}{\del C^1_{\b_1}} \cdots \fr{\del}{\del C^{11}_{\b_{11}}}\prod_{I=1}^{11} (\fr{\del}{\del \lambda^\d}\fr{\del}{\del C^I_{\d}})^{K_I}\prod_{P=1}^{10}\prod_{R=1}^g (\fr{\del}{\del B^P_{pq}} \fr{\del}{\del N_R^{pq}})^{L_{P,R}} \prod_{R=1}^g (\fr{\del}{\del J_R})^{M_R} h(\lambda,N_R,J_R,C,B),
\]
for some proportionality constant $c$.

In the sequel we will use the following notation
\bea \label{eq:defetR}
(\e TR)^{\a_1\cdots \a_{11}}_{\b_1\cdots \b_{11}m_1n_1\cdots m_{10}n_{10}}&\equiv& (\e T)^{((\a_1\a_2\a_3}_{\b_1\cdots \b_{11}}R^{\a_4\cdots \a_{11}))}_{m_1n_1\cdots m_{10}n_{10}},
\\
(TR)^{\a_1\cdots \a_{11}\b_{12}\cdots \b_{16}}_{m_1n_1\cdots m_{10}n_{10}}&\equiv& T^{\b_{12}\cdots \b_{16}((\a_1\a_2\a_3} R^{\a_4\cdots \a_{11}))}_{m_1n_1\cdots m_{10}n_{10}}.
\eea

\section{Proof of decoupling of $Q$ exact states} \label{sec:proof}
Decoupling of unphysical states in the minimal pure spinor formalism would follow, if
all insertions were $Q$ closed.
As discussed in \cite{new}, however, the $Y$'s are not BRST closed inside correlators.
More specifically the PCO's for the $\lambda$ zero modes, denoted by $Y_C$, are not closed:
\be
Q Y_C=Q (C\q)\d(C\lambda)=(C\lambda)\d(C\lambda).
\ee
This vanishes in a distributional sense but it does not necessarily vanish if there are insertions containing factors of $\fr{1}{C\lambda}$. In the minimal pure spinor formalism the pure spinor zero mode measure $[d\lambda]$ contains a factor $\fr{1}{(\lambda^+)^3}$. Therefore one cannot conclude $QY_C=0$. In fact in \cite{new} some amplitudes with $Q$ exact states were computed in the formulation without an integral over $C$. These did not vanish, hence we can conclude $Q(Y_{C^1}\cdots Y_{C^{11}})\neq 0$. Nevertheless, we were able to show
decoupling of $Q$ exact states at tree level in the formulation with an integral over $C$. In next subsection
we review the tree-level argument in a form that generalizes to the higher loops
and show that $Q$ exact states decouple to all orders. A crucial role in this proof is played by symmetry of the insertions that follows from the particular form of the picture raising operators, $Z_B$. We will first present the proof for one-loop amplitudes, followed by a proof of decoupling valid at any genus.

\subsection{Tree-level amplitudes} \label{sec:tree}
After integrating out the non-zero modes every tree-level amplitude assumes the form
\be \label{eq:treezero}
\cA=\int [d\lambda][dC]d^{16}\q \lambda^\a\lambda^\b\lambda^\g f_{\a\b\g}(\q,a,k)\q^{\b_1}\cdots \q^{\b_{11}}C^1_{\b_1}\cdots C^{11}_{\b_{11}}\d(C^1\lambda)\cdots \d(C^{11}\lambda),
\ee
where $a$ denotes all polarizations and $k$ denotes all momenta.
Note that we have assumed that integration
over the non-zero modes does not affect the factor
of $Y_{C^1} \cdots Y_{C^{11}}$. This can be justified
either by writing $Y_C$ as a function of only zero modes
or by inserting the factor of $(Y_C)^{11}$ at $z=\infty$ on the worldsheet.
The three factors of $\lambda$ originate from the three
unintegrated vertex operators and the factors of $\theta$, $C$ and $\delta(C \lambda)$ from the eleven picture changing operators $Y_C$.
In order to evaluate \eqref{eq:treezero} first note that only terms with five $\q$'s can contribute:
\be \label{eq:tz2}
\cA= \int [d\lambda][dC]d^{16}\q \lambda^\a\lambda^\b\lambda^\g f^{(5)}_{\a\b\g\b_{12}\cdots\b_{16}}(a,k)\q^{\b_1}\cdots \q^{\b_{16}}C^1_{\b_1}\cdots C_{\b_{11}}\d(C^1\lambda)\cdots \d(C^{11}\lambda),
\ee
We will show now that the integration is a projection on the scalar in $f^{(5)}_{\a\b\g\b_{12}\cdots\b_{16}}(a,k)$. To this end we write the tensor product $(\lambda)^3(\q)^5$ in terms of its irreducible representations:
\be \label{eq:tensorexp}
\lambda^\a\lambda^\b\lambda^\g\q^{\b_{12}}\cdots \q^{\b_{16}}=T^{\a\b\g\b_{12}\cdots \b_{16}}T_{\a'\b'\g'\b'_{12}\cdots \b'_{16}}\lambda^{\a'}\lambda^{\b'}\lambda^{\g'}\q^{\b'_{12}}\cdots \q^{\b'_{16}}+
\ee
\[
(T_1)^{\a\b\g\b_{12}\cdots \b_{16}[mn]}(T_1)_{\a'\b'\g'\b'_{12}\cdots \b'_{16}[mn]}\lambda^{\a'}\lambda^{\b'}\lambda^{\g'}\q^{\b'_{12}}\cdots \q^{\b'_{16}}+
\]
\[
\sum_{i\geq 2} (T_i)^{\a\b\g\b_{12}\cdots \b_{16}x_i}(T_i)_{\a'\b'\g'\b'_{12}\cdots \b'_{16}x_i}\lambda^{\a'}\lambda^{\b'}\lambda^{\g'}\q^{\b'_{12}}\cdots \q^{\b'_{16}},
\]
where $x_i$ are the indices representing the representation. To obtain the above expansion one first needs to compute the tensor product ${\rm Gam}^3 {\bf 16} \otimes {\rm Asym}^5 {\bf 16}$. This contains one scalar, hence there is only one invariant tensor with the indices and symmetries of $T$ (an explicit realization is specified in \eqref{eq:deft}). One also finds there is one ${\bf 45}$ in the tensor product, hence the second line. The sum in the last line runs over all the other irreps in the tensor product, each one has an invariant tensor $(T_i)$ associated to it. Furthermore all the $(T_i)$'s satisfy
\be \label{eq:ortho}
T^{\a\b\g\b_{12}\cdots \b_{16}}(T_i)_{\a\b\g\b_{12}\cdots \b_{16}x_i}=0.
\ee
This can be proven by contracting both sides of \eqref{eq:tensorexp} with $T_{\a\b\g\b_{12}\cdots \b_{16}}$. The integrations in \eqref{eq:tz2} can be evaluated by Lorentz invariance:
\[
\left(\int d^{16} \q \q^{\b_1}\cdots \q^{\b_{16}}\right)\left(\int [d\lambda][dC] \lambda^\a\lambda^\b\lambda^\g C^1_{\b_1}\cdots C_{\b_{11}}\d(C^1\lambda)\cdots \d(C^{11}\lambda)\right)=
\]
\be \label{eq:intev}
\e^{\b_1\cdots \b_{16}}(\e T)^{\a\b\g}_{\b_1\cdots \b_{11}}=T^{\a\b\g\b_{12}\cdots \b_{16}}
\ee
After using \eqref{eq:ortho} one sees all the non scalar terms in \eqref{eq:tensorexp} are annihilated by the integration. It is therefore a projection on the scalar as claimed. The final expression for the amplitude becomes
\be
\cA=T^{\a\b\g\b_{12}\cdots \b_{16}}f^{(5)}_{\a\b\g\b_{12}\cdots \b_{16}}(a,k).
\ee

\subsubsection{Decoupling of $Q$ exact states at tree level} \label{sec:dectree}
After integrating out the non-zero modes, the amplitude containing a $Q$-exact state becomes,
\be \label{eq:qextr}
\int [d\lambda]d^{16}\q (Q \W(\lambda,\q,a,k)) \q^{\b_1}\cdots \q^{\b_{11}}C^1_{\b_1}\cdots C_{\b_{11}}\d(C^1\lambda)\cdots \d(C^{11}\lambda),
\ee
for some $\W$, where all fields are zero modes. Our task now is to show that this integral
vanishes for any $\W$.

Since only the terms with five $\q$'s and three $\lambda$'s in $Q\W$ contribute, we focus on terms in $\W$ with two $\lambda$'s and six $\q$'s. The upshot of the proof is that no Lorentz scalar can be constructed from two $\lambda$'s and six $\q$'s. Therefore there will be no scalar in $Q(\lambda)^2(\q)^6$ and since the integration projects on the scalar the amplitude vanishes. In order to make this argument precise let us write:
\be
\W|_{(\lambda)^2(\q)^6}=\lambda^{\a}\lambda^{\b}\q^{\b_1}\cdots \q^{\b_6}\tilde{f}_{\a\b\b_1\cdots \b_6}(a,k)
\ee
for some $\tilde{f}$. The next step is writing the tensor product $(\lambda)^2(\q)^6$ in terms of its irreducible representations:
\be
\W|_{(\lambda)^2(\q)^6}=\tilde{f}_{\a\b\b_1\cdots \b_6}(a,k)\left( \sum_i (\tilde{T}_i)^{\a\b\b_1\cdots \b_{6}y_i} (\tilde{T}_i)_{\a'\b'\b'_1\cdots \b'_{6}y_i}\lambda^{\a'}\lambda^{\b'}\q^{\b'_1}\cdots \q^{\b'_6}\right).
\ee
In the above formula it is important to note that there are no scalars in the tensor product of two pure spinors and six fermionic spinors. This is reflected by the fact that $y_i$ represents (a positive number of) indices for every $i$. We now perform the $Q$ transformation:
\be
Q\W|_{(\lambda)^2(\q)^6}=\tilde{f}_{\a\b\b_1\cdots \b_6}(a,k)\left( \sum_i (\tilde{T}_i)^{\a\b\b_1\cdots \b_{6}y_i} (\tilde{T}_i)_{\a'\b'[\g'\b'_2\cdots \b'_{6}]y_i}\lambda^{\a'}\lambda^{\b'}\lambda^{\g'}\q^{\b'_2}\cdots \q^{\b'_6}\right).
\ee
After invoking \eqref{eq:intev} we find
\be
\int [d\lambda]d^{16}\q \left(Q\W|_{(\lambda)^2(\q)^6}\right) \q^{\b_1}\cdots \q^{\b_{11}}C^1_{\b_1}\cdots C_{\b_{11}}\d(C^1\lambda)\cdots \d(C^{11}\lambda) =
\ee
\[
\tilde{f}_{\a\b\b_1\cdots \b_6}(a,k) \sum_i (\tilde{T}_i)^{\a\b\b_1\cdots \b_{6}y_i} (\tilde{T}_i)_{\a'\b'[\g'\b'_2\cdots \b'_{6}]y_i}T^{\a'\b'[\g'\b'_2\cdots \b'_6]}=0
\]
This vanishes because
\be \label{l2q6}
T^{\a'\b'[\g'\b'_2\cdots \b'_6]}=0,
\ee
which follows from the statement that there are no scalars in $(\lambda)^2(\q)^6$. This concludes the proof that \eqref{eq:qextr} vanishes.

\subsection{Higher-loop amplitudes}
In order to prove decouling of unphysical states at higher-loop amplitudes one can take similar steps to the tree-level case. This means that one first reduces the amplitude to a zero mode integral, which is effectively a projection onto a scalar and then one shows there is no scalar when one started with a $Q$ exact state. In the higher-loop case we need an additional ingredient for the second step which is a symmetry possessed by the integrand of the functional integral.

\subsubsection*{Additional symmetry}
The amplitude prescription contains products of PCO's $Z_B$ and $Z_J$.
The main observation is that
\be
Z_B Z_J= B_{mn} \lambda\g^{mn}d ~ \d (B_{mn} N^{mn}) (\lambda d) \delta(J)
\ee
is invariant under
\be \label{eq:btrans}
\d B_{mn}=(\lambda\g_{[m})_{\a}f _{n]}^{\a}.
\ee
where $f^{n\a}$ are constants.
This transformation acts on the $B_{mn}N^{mn}$ and $B_{mn}\lambda\g^{mn}d$ as,
\bea
\d B_{mn}N^{mn}&=&(\lambda\g_m)_{\a}f^\a_n (\lambda\g^{mn}w) = (\lambda\g^n f_n) (\lambda w),
\\
\d B_{mn}(\lambda\g^{mn}d)&=& (\lambda\g_m)_{\a}f^\a_n (\lambda\g^{mn}d) = (\lambda\g^n f_n) (\lambda d).
\eea
Since all these transformation contain either $(\lambda w)$ or $(\lambda d)$ and $Z_J$ contains both $\d(\lambda w)$ and
$\lambda d$:
\be
\d (Z_B Z_J) =0.
\ee
Now recall that at genus $g$, $3 g-3$ $B$'s (one at genus one) enter via the $b$ ghost.
We take these $B$'s to be inert. The remaining $7 g + 3$ $B$'s (9 at genus one) are
taken to transform as in \eqref{eq:btrans} . Note that at one loop,
the factor of $(Z_{B})^{9} Z_J$ is placed at a single
point on the worldsheet. At two-loop order, the additional factor
of
$(Z_{B})^7 Z_J$ is placed at a second point on the worldsheet. And
at each additional loop order, one places the new factor of
$(Z_{B})^7 Z_J$ at a $g^{th}$ point on the worldsheet. With this choice,
\eqref{eq:btrans} is an invariance of the theory for $7g+3$
$B$'s and
the amplitudes must respect this symmetry.

One can understand the origin of this symmetry by going back to the first principles
derivation of the amplitude prescription in \cite{Hoogeveen:2007tu}. As shown there,
PCO insertions arise from gauge fixing the invariance due to pure spinor zero modes.
In particular, this leads to the insertion in the path integral
of the expression
\be
\exp \left( \pi_\a \lambda^\a + \tilde{\pi}_\a \theta^\a
+ \sum_{I=1}^g \left(-\pi_{mn}^{I} \frac{1}{2} d^I \g^{mn} \lambda +
\tilde{\pi}_{mn}^{I} N^{mnI} - \pi^I d_\a^I \lambda^\a
+ \tilde{\pi}^I J^I \right) \right),
\ee
where $\pi_\a, \tilde{\pi}_\a, \pi^{mnI}, \tilde{\pi}^{mnI}, \pi^I, \tilde{\pi}^I$ are
the BRST auxiliary fields, $g$ is the genus and $I$ counts the $g$ zero modes of worldsheet
vectors. This expression is invariant under
\bea \label{new_sym}
\d \pi^I_{mn}&=&(\lambda\g_{[m})_{\a}f _{n]}^{I\a}, \qquad  \delta \pi^I = - \frac{1}{2} (\lambda\g^n f_n^I) \\
\d \tilde{\pi}^I_{mn}&=&(\lambda\g_{[m})_{\a}\tilde{f} _{n]}^{I\a}, \qquad
\delta \tilde{\pi}^I = - (\lambda\g^n \tilde{f}_n^I)
\eea
This symmetry implies that out of the 45 components of each $\pi_{mn}^I$ and $\tilde{\pi}_{mn}^I$ only 10 are independent, as it should be since the number of BRST auxiliary fields
should be equal to the number of gauge fixing conditions.

In the non-minimal formalism, $\pi_{mn}^I$ and $\tilde{\pi}_{mn}^I$
are identified with $\bar{N}^{mnI}$ and $S_{mn}^I$ and these
automatically have the correct number of independent components because they
are constructed using the non-minimal pure spinor variables. On the other hand,
to arrive at the minimal formulation we set
\be
\pi_{mn}^{I} = p^{jI} B_{mn}^{jI}, \qquad \tilde{\pi}_{mn}^{I} =
\tilde{p}^{jI} B_{mn}^{jI}, \qquad j=1,\ldots, 10
\ee
and integrate over $p^{jI}, \tilde{p}^{jI}$. This leads to the $Z_B$ insertions.
In this case, there is a remnant of the symmetry (\ref{new_sym}), which is
\eqref{eq:btrans} for all $B$. This suggests that the amplitudes is also invariant,
if we transform the $(3 g -3)$ (one when $g=1$) factors of $B$ involved in the $b$ insertions\footnote{Recall that $(3 g-3)$ (one when $g=1$) of the $Z_B$ factors are absorbed into the
$b$-insertions.}, but we will
not prove or use this here.

\subsubsection{One-loop amplitudes}
After integrating out all non-zero modes, as well as the
$d_\alpha$ zero modes, every one-loop amplitude can be written as
\be \label{eq:1lzm}
\int [d\lambda][dN][dC][dB]d^{16}\q \lambda^{\a_1}\cdots \lambda^{\a_{11}} B^1_{m_1n_1}\cdots B^{10}_{m_{10}n_{10}} f_{\a_1\cdots \a_{11}}^{m_1n_1\cdots m_{10}n_{10}}(\q,a,k)
\ee
\[
\q^{\b_1}\cdots \q^{\b_{11}}C^1_{\b_1}\cdots C^{11}_{\b_{11}}\d(C^1\lambda)\cdots \d(C^{11}\lambda)\d(B^1N)\cdots \d(B^{10}N)\d(J),
\]
where all fields are zero modes and the integrand is invariant under the $B$ transformation \eqref{eq:btrans}.
As in the tree amplitude, we are assuming that integration over
the non-zero modes does not affect the $(Y_{C})^{11}$ factor since
this factor can be written in terms of only zero modes.
In this expression, eleven factors of $\lambda$ originate as follows: one from the unintegrated vertex operator, one from $Z_J$
and nine from the nine factors of $Z_B$. In general the zero mode integral can contain additional
factors of the Lorentz currents $N$, higher
powers of $B$ and higher derivatives of $\delta(B N)$.
These additional factors can be put into the form of
\eqref{eq:1lzm} by integrating by parts using that
$N^{pq} B_{mn} \partial \delta (BN) = - \delta_m^{[p} \delta_n^{q]} \delta (BN)$.

One can show that the integral in \eqref{eq:1lzm} is also a projection on a scalar. To see this first note that there is one scalar in ${\rm Gam}^{11} {\bf 16}\otimes {\rm Asym}^5 {\bf 16}\otimes {\rm Asym}^{10}{\bf 45}$. This implies one can write
\[
\lambda^{\a_1}\cdots \lambda^{\a_{11}}\q^{\b_{12}}\cdots \q^{\b_{16}}B^1_{m_1n_1}\cdots B^{10}_{m_{10}n_{10}}= (TR)^{\a_1\cdots\a_{11}\b_{12}\cdots \b_{16}}_{m_1n_1\cdots m_{10}n_{10}}\left((TR)(\lambda)^{11}(\q)^5(B)^{10}\right)+
\]
\be
\sum_i (S_i)^{\a_1\cdots\a_{11}\b_{12}\cdots \b_{16}}_{m_1n_1\cdots m_{10}n_{10}x_i}\left(S_i(\lambda)^{11}(\q)^5(B)^{10}\right)^{x_i},
\ee
where the notation $\left((TR)(\lambda)^{11}(\q)^5(B)^{10}\right)$ means that all indices of $(TR)$ have been contracted with those of $\lambda,\q$ and $B$ and $\left(S_i(\lambda)^{11}(\q)^5(B)^{10}\right)^{x_i}$ denotes an object that has $x_i$ as its only free index and which
transforms in some non-scalar representation. Similar to the tree-level case the invariant tensors $S_i$ satisfy
\be \label{eq:ortho1l}
\left ((RT) (S_i) \right)^{x_i}=0.
\ee
Note that since $B$ is not a covariant tensor this is not the decomposition of a Lorentz invariant object into a lot of Lorentz invariant terms like \eqref{eq:tensorexp}. However this does not matter, the point of performing this expansion is that all the non scalar terms vanish due to the integration. The last point follows from \eqref{eq:ortho1l} and
\be \label{eq:ert}
\int [d\lambda][dC][dB][dN] \lambda^{\a_1}\cdots \lambda^{\a_{11}}B^1_{m_1n_1}\cdots B^{10}_{m_{10}n_{10}}C^1_{\b_1}\cdots C^{11}_{\b_{11}}
\ee
\[
\d(C^1\lambda)\cdots \d(C^{11}\lambda)\d(B^1N)\cdots \d(B^{10}N)\d(J)= (\e TR)^{\a_1\cdots \a_{11}}_{\b_1\cdots \b_{11}m_1n_1\cdots m_{10}n_{10}},
\]
which is also a consequence of the fact there is only one Lorentz scalar in ${\rm Gam}^{11} {\bf 16}\otimes {\rm Asym}^5 {\bf 16}\otimes {\rm Asym}^{10}{\bf 45}$.

\subsubsection*{Decoupling of $Q$ exact states}
We will show that if
\be
\lambda^{\a_1}\cdots \lambda^{\a_{11}} B^1_{m_1n_1}\cdots B^{10}_{m_{10}n_{10}} f_{\a_1\cdots \a_{11}}^{m_1n_1\cdots m_{10}n_{10}}(\q,a,k)
\ee
can be written as $Q \W$ where $\W$ is invariant under the $B$ transformation then \eqref{eq:1lzm} vanishes.

Note $\W$ must contain ten $\lambda$'s, six $\q$'s and ten $B$'s. There are two scalars in ${\rm Gam}^{10}{\bf 16} \otimes {\rm Asym}^6 {\bf 16} \otimes {\rm Asym}^{10} {\bf 45}$. Since ${\rm Gam}^{11} {\bf 16}\otimes {\rm Asym}^5 {\bf 16}\otimes {\rm Asym}^{10}{\bf 45}$ contains only a single scalar and $Q$ maps scalars to scalars, there is a basis of invariant tensors such that one of the scalars is annihilated by the BRST operator
and the other one, call it $\Omega_1$, has non-zero BRST variation, $Q \Omega_1 \neq 0$.
This scalar is\footnote{Another possible candidate, $\left( T (\lambda)^2 (\q)^6 \right) \left( R (B)^{10} (\lambda)^8\right)$, vanishes identically because of (\ref{l2q6}).}
\be \label{o1}
\Omega_1 = \left( T (\lambda)^3 (\q)^5 \right) \left( R (B)^{10} (\lambda)^7 (\q)^1\right).
\ee
Here $(R (B)^{10} (\lambda)^7 (\q)^1)$ denotes the unique scalar obtained by contracting all indices of the
objects involved. The state $Q\Omega_1$
is a candidate BRST exact state that may not decouple.
The scalar $\Omega_1$ however is not invariant under the
transformation \eqref{eq:btrans} for 9 of the 10 $B$'s.
In fact, one can show that $\Omega_1$ is invariant under the
transformation \eqref{eq:btrans} for only
6 of the 10 $B$'s. To see this, note that
$\left( R(B)^{10} (\lambda)^7 (\theta)^1 \right)$ can be expressed
as
\be
(\lambda\gamma^{m_1 \cdots m_5}\lambda) (\lambda\gamma^{m_6 \cdots m_{10}}\lambda) (\lambda\gamma^{m_{11} \cdots m_{15}}\lambda) (\lambda\gamma^{m_{16} \cdots m_{20}}\theta)
\ee
contracted with the 20 vector indices of $(B)^{10}$.
If both indices of $B_{pq}$ are contracted with  $m_1 \cdots m_{15}$,
then $\Omega_1$ is invariant under the transformation
\eqref{eq:btrans} for that $B$ since
$(\lambda \g^{mn_1 \cdots n_4} \lambda) (\lambda \g_m)_\a =0$.
However, if at least one index of
$B_{pq}$ is contracted with $m_{16} ... m_{20}$, then $\Omega_1$
is not invariant under the transformation
\eqref{eq:btrans} for that $B$. Using the 
definition of $R^{\alpha_1 ... \alpha_8}_{m_1 ... m_{20}}$,
one finds there are four $B$'s whose
indices are contracted with $m_{16} ... m_{20}$,
so $\Omega_1$ is invariant under the
transformation \eqref{eq:btrans} for 6 of the 10 $B$'s.

But since the gauge parameter must be
invariant under
\eqref{eq:btrans} 
for 9 of the 10 $B$'s, there is no way to generate $\Omega_1$ as
a possible gauge parameter.
We thus conclude that if it is $Q$ exact and invariant under the $B$ transformation,
\be
f_{\a_1\cdots \a_{11}}^{m_1n_1\cdots m_{10}n_{10}}(\q,a,k)\lambda^{\a_1}\cdots \lambda^{\a_{11}}B^1_{m_1n_1}\cdots B^{10}_{m_{10}n_{10}}
\ee
does not contain any scalars constructed from eleven $\lambda$'s, five $\q$'s and ten $B$'s. Since the integration projects on the (single) scalar the total zero mode integral vanishes. The precise argument is analogous to the steps in section \ref{sec:dectree}.

\subsubsection{Higher-loop amplitudes}
The argument for $g>1$ is exactly analogous.
After integrating out all non-zero modes,
as well as the zero modes of $d_\alpha$,
every $g>1$ loop amplitude can be written as
\bea \label{eq:glzm}
&&\int d^{16}\q [d\lambda][dC] \lambda^{\a_1}\lambda^{\a_2} \lambda^{\a_3} \q^{\b_1}\cdots \q^{\b_{11}}C^1_{\b_1}\cdots C^{11}_{\b_{11}}\d(C^1\lambda)\cdots \d(C^{11}\lambda)  \nonumber \\
&& \prod_{I=1}^g
 \left([dB^I] [dN^I] \lambda^{\a^I_4}\cdots \lambda^{\a^I_{11}} B^{1 I}_{m^I_1n^I_1}\cdots B^{10 I}_{m^I_{10}n^I_{10}}
 \d(B^{1 I} N)\cdots \d(B^{10 I}N)\d(J^I) \right) \nonumber \\
&& \qquad f_{\a_1 \a_1 \a_3 \a_4^1 \cdots \a^g_{8}}^{m^1_1n^1_1\cdots m^g_{10}n^g_{10}}(\q,a,k)
\eea
where all fields are zero modes and the integrand is invariant under the $B$ transformation \eqref{eq:btrans}.
Now the factors $\lambda$ originate from the $(7 g + 3)$ factors of $Z_B$ and the $g$ factors of $Z_J$.
Additional factors of $N$, $B$ and derivatives of $\delta(B N)$ can be removed as in the one-loop case.

In this case the analogue of \eqref{eq:ert} is
\bea
&&\int [d\lambda][dC] \lambda^{\a_1}\lambda^{\a_2} \lambda^{\a_3} C^1_{\b_1}\cdots C^{11}_{\b_{11}}\d(C^1\lambda)\cdots \d(C^{11}\lambda)  ,\nonumber \\
&& \prod_{I=1}^g
 \left([dB^I] [dN^I] \lambda^{\a^I_4}\cdots \lambda^{\a^I_{11}} B^{1 I}_{m^I_1n^I_1}\cdots B^{10 I}_{m^I_{10}n^I_{10}}
 \d(B^{1 I} N)\cdots \d(B^{10 I}N)\d(J^I) \right) \nonumber \\
&& = (\e T R^g)^{\a_1 \a_2 \a_3 \a_4^1 \cdots \a_{11}^g}_{\b_1 \cdots \b_{11} m_1^1 n_1^1 \cdots m_{10}^g n_{10}^g}
\eea
where $(\e T R^g)$ is the generalization of (\ref{eq:defetR}) involving $g$ factors of $R$.

There are $g$ candidate BRST exact states that may not decouple, which
are the analogs of
\eqref{o1} and are given by
\be
\Omega_J =
 \left( T (\lambda)^3 (\q)^5 \right)
 \prod_{I=1}^{J-1} \left( R (B^I )^{10} (\lambda)^8 \right) ~~
\left( R (B^J )^{10} (\lambda)^7 (\q)^1\right)~~
\prod_{I=J+1}^g \left( R (B^I )^{10} (\lambda)^8 \right)
\ee
where $B^I$ denotes the $B$'s associated with the $I^{th}$ zero mode.
As in the one-loop case, the term
$\left( R (B^J )^{10} (\lambda)^7 (\q)^1\right)$ is at most invariant
under 6 of the 10 $B^J$ transformations. But invariance under
\eqref{eq:btrans} requires invariance under 7 of the 10 $B^J$ transformations.

So we conclude that unphysical states decouple to all orders in $g$.

\section{Conclusion} \label{conclu}
We presented in this paper a proof of decoupling of unphysical states in the minimal pure spinor formalism 
to all loop order. We were able to prove this despite the fact that not all insertions in the 
path integral are $Q$ closed. More specifically our argument did not involve integrating $Q$ by parts. The two main ingredients were the presence of the $B$ symmetry  
and the fact that the zero mode integrals act as projectors on a scalar.


As is discussed in \cite{new}, the amplitudes in the prescription of \cite{Berkovits:2004px} without an integral over $C$ are actually singular and the distributional relations do not hold inside correlators.
The singularities in the amplitudes are likely to reflect the fact that the gauge choice for the gauge invariances due to zero modes
implicit in the prescription of \cite{Berkovits:2004px} is singular. Obtaining a prescription corresponding to a non-singular gauge choice
may require incorporating global issues, in particular taking into account  all patches in the pure spinor space\footnote{We thank
Nikita Nekrasov for this suggestion.}, see section 6 of \cite{new} for further discussion, and such a prescription is
currently under investigation \cite{progress}. We anticipate that such a prescription will also lead
to decoupling of BRST exact states, without integrating over $C$ and $B$.









\section*{Acknowledgments}

We would like to thank Carlos Mafra and Nikita Nekrasov for discussions.
KS is supported in part by NWO. KS and NB would like to thank KITP for
hospitality during initial stages of this work. KS would like to thank the Aspen Center of Physics
for hospitality during the final stages of this work.
This work was supported in part by the National Science Foundation under
Grant No. NSF PHY05-51164. The research of NB was also partially
supported by CNPq grant 300256/94-9 and FAPESP grant 04/11426-0.

\appendix

\section{Chain of operators for $b$ ghost} \label{sec:appco}

\bea \label{eq:bB0}
{b_B}_0&=&\hp G\g^{mn}dB_{mn}-\hp H^{\a\b}(\g^p\g^{mn})_{\a\b}\Pi_pB_{mn}+\\ \nn
&&\hp K^{\a\b\g}(\g^p\g^{mn})_{\b\g}(\g_p\del \q)_{\a}B_{mn}+\hp S^{\a\b\g}(\g^p\g^{mn})_{\b\g}(\g^p\del \lambda)_{\a}B_{mn},\\
{b_B}_1&=&\qu H^{\a\b} (Bd)_{\a}(Bd)_{\b}+\\ \nn
&&\qu K^{\a\b\g}(\g^p\g^{mn})_{\b\g}(Bd)_{\a}\Pi_pB_{mn}+\qu K^{\a\b\g}(\g^p\g^{mn})_{\a[\b}(Bd)_{\g]}\Pi_pB_{mn}+\\ \nn
&&\qu L^{\a\b\g\d}[((\g^p\g^{mn})_{\g\d}(Bd)_{[\a}(\g_p\del \q)_{\b]}-(\g^p\g^{mn})_{\b[\g}(Bd)_{\d]}(\g_p\del \q)_{\a})B_{mn}-\\ \nn
&&((\g^s\g^{rq})_{\a[\b}(\g^p\g^{mn})_{\g]\d}+(\g^s\g^{rq})_{\a\d}(\g^p\g^{mn})_{\b\g})\Pi_pB_{mn}\Pi_sB_{qr}],\\
{b_B}_2&=&-\fr{1}{8}K^{\a\b\g}(Bd)_{\a}(Bd)_{\b}(Bd)_{\g}-\fr{1}{8}L^{\a\b\g\d}((\g^p\g^{mn})_{\g\d}(Bd)_{\b}(Bd)_{\a}+\\ \nn
&&(\g^p\g^{mn})_{\b[\g}(Bd)_{\d]}(Bd)_{\a}+\hp (\g^p\g^{mn})_{\a[\d}(Bd)_{\g}(Bd)_{\b]})\Pi_pB_{mn},\\
{b_B}_3&=&-\fr{1}{16}L^{\a\b\g\d}(Bd)_{\a}(Bd)_{\b}(Bd)_{\g}(Bd)_{\d}, \label{eq:bB3}
\eea
where $(Bd)_{\a}\equiv B_{mn}(\g^{mn}d)_{\a}$.
The explicit form of the
tensors $G^\a,H^{\a\b},K^{\a\b\g},L^{\a\b\g\d}$ can be found, for example, in
section 3 of \cite{Oda:2007ak}. They do not contain any $B$ tensors.

\providecommand{\href}[2]{#2}\begingroup\raggedright\endgroup


\end{document}